\documentclass[aps,prl,twocolumn]{revtex4-1}
\usepackage{graphicx}
\usepackage{hyperref}
\usepackage{amsmath}
\usepackage{amssymb}

\newcommand{\beq}{\begin{eqnarray}}
\newcommand{\eeq}{\end{eqnarray}}
\def\simlt{\stackrel{<}{{}_\sim}}
\def\simgt{\stackrel{>}{{}_\sim}}

\begin{document}

\title{A Simple Recipe for the 111 and 128 GeV Lines}

\author{JiJi Fan}

\author{Matthew Reece}
\affiliation{Department of Physics, Harvard University, Cambridge, MA, 02138}

\begin{abstract}
Recently evidence for gamma ray lines at energies of approximately 111 and 128 GeV has been found in Fermi-LAT data from the center of the galaxy and from unassociated point sources. Many explanations in terms of dark matter particle pairs annihilating to $\gamma\gamma$ and $\gamma Z$ have been suggested, but these typically require very large couplings or mysterious coincidences in the masses of several new particles to fit the signal strength. We propose a simple novel explanation in which dark matter is part of a multiplet of new states which all have mass near 260 GeV as a result of symmetry. Two dark matter particles annihilate to a pair of neutral particles in this multiplet which subsequently decay to $\gamma\gamma$ and $\gamma Z$. For example, one may have a triplet of pseudo-Nambu-Goldstone bosons, $\pi^h_\pm$ and $\pi^h_0$, where $\pi^h_\pm$ are stabilized by their charge under a new U(1) symmetry and the slightly lighter neutral state $\pi^h_0$ decays to $\gamma\gamma$ and $\gamma Z$. The symmetry structure of such a model explains the near degeneracy in masses needed for the resulting photons to have a line-like shape and the large observed flux. The tunable lifetime of the neutral state allows such models to go unseen at direct detection or collider experiments that can constrain most other explanations. However, nucleosynthesis constraints on the $\pi^h_0$ lifetime fix a minimum necessary coupling between the new multiplet and the Standard Model. The spectrum is predicted to be not a line but a box with a width of order a few GeV, smaller than but on the order of the Fermi-LAT resolution.
\end{abstract}


\maketitle

{\bf Introduction:} Dark matter makes up 80\% of the matter in our universe, but its nature continues to be elusive. A number of independent lines of evidence offer a persuasive picture of dark matter's existence and gravitational interactions, but it is unclear whether it has interactions that are stronger than gravity with known Standard Model particles. Recently, a striking observation has been made of monochromatic gamma ray emission near the center of the galaxy~\cite{Bringmann:2012vr,Weniger:2012tx}, with energy about 128 GeV. Subsequent studies~\cite{Rajaraman:2012db,Su:2012ft} have shown that there may be a second line with an energy of about 111 GeV (see also~\cite{Boyarsky:2012ca}), and that both lines also show up in unassociated sources in the Fermi-LAT catalogue~\cite{Su:2012zg}. This is suggestive of dark matter annihilating to $\gamma\gamma$ and $\gamma Z$, with the unassociated sources as potential dark matter subhalos within the Miky Way. However, there remain uncertainties about whether the line could be due to a systematic error, especially due to troubling hints of a signal in the Earth limb~\cite{Su:2012ft, Bloom:2013mwa}.

\begin{figure}[!h]
\includegraphics[width=0.7\columnwidth]{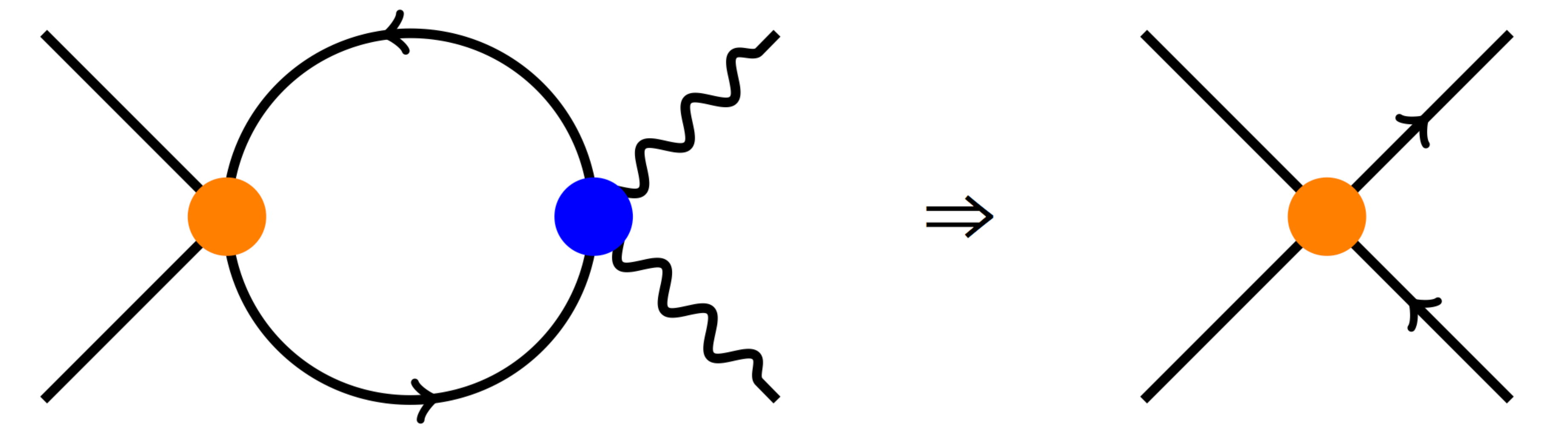}
\caption{A model of ${\rm DM}+{\rm DM} \to \gamma+\gamma$ often implies the existence of a tree-level annihilation, by cutting the loop.\label{fig:cutloop}}
\end{figure}

Because dark matter charge is constrained to be tiny~\cite{McDermott:2010pa, Cline:2012is}, a model in which two dark matter particles annihilate to two photons will generally rely on annihilation through a loop of charged particles. As illustrated in Figure~\ref{fig:cutloop}, this will imply the existence of a tree-level annihilation process to charged particles (whenever they are light enough to be kinematically accessible). These charged particles can radiate photons and frequently decay to showers of hadrons that can in turn decay to further photons. This would appear as a continuum spectrum of gamma rays that have not yet been seen in Fermi-LAT data, ruling out many models fitting the lines, including MSSM neutralinos~\cite{Buchmuller:2012rc,Cohen:2012me,Cholis:2012fb} (except for a tuned case involving internal bremsstrahlung~\cite{Bringmann:2012vr,Bringmann:2012ez}).

Estimates of the strength of the line vary from about 1.3 to 5.1 $\times 10^{-27}~{\rm cm}^3/{\rm s}$~\cite{Weniger:2012tx,Tempel:2012ey}, and depend to some extent on assumptions about the halo properties. For the simplest cases of DM annihilating through a loop, this requires rather large couplings, even allowing for numerical enhancements from coincidences in the mass of the DM and the charged particle in the loop~\cite{Cline:2012nw,Choi:2012ap,Buckley:2012ws}. (Similar remarks apply to UV completions of MiDM/RayDM~\cite{Weiner:2012cb,Weiner:2012gm}.) Another possible source of enhancement is from $s$-channel exchange of a pseudoscalar~\cite{Buckley:2012ws,Lee:2012bq,Das:2012ys,Tulin:2012uq} or (for $\gamma Z$ without $\gamma\gamma$) vector~\cite{Dudas:2012pb}, but this again requires a tuning of the mass in the propagator for an enhancement. (Another interesting model that predicts this topology is Goldstone fermion dark matter~\cite{Bellazzini:2011et}.) These models could be probed at colliders~\cite{Weiner:2012cb} or in direct detection experiments~\cite{Frandsen:2012db}.

\begin{figure}[!h]
\includegraphics[width=0.4\columnwidth]{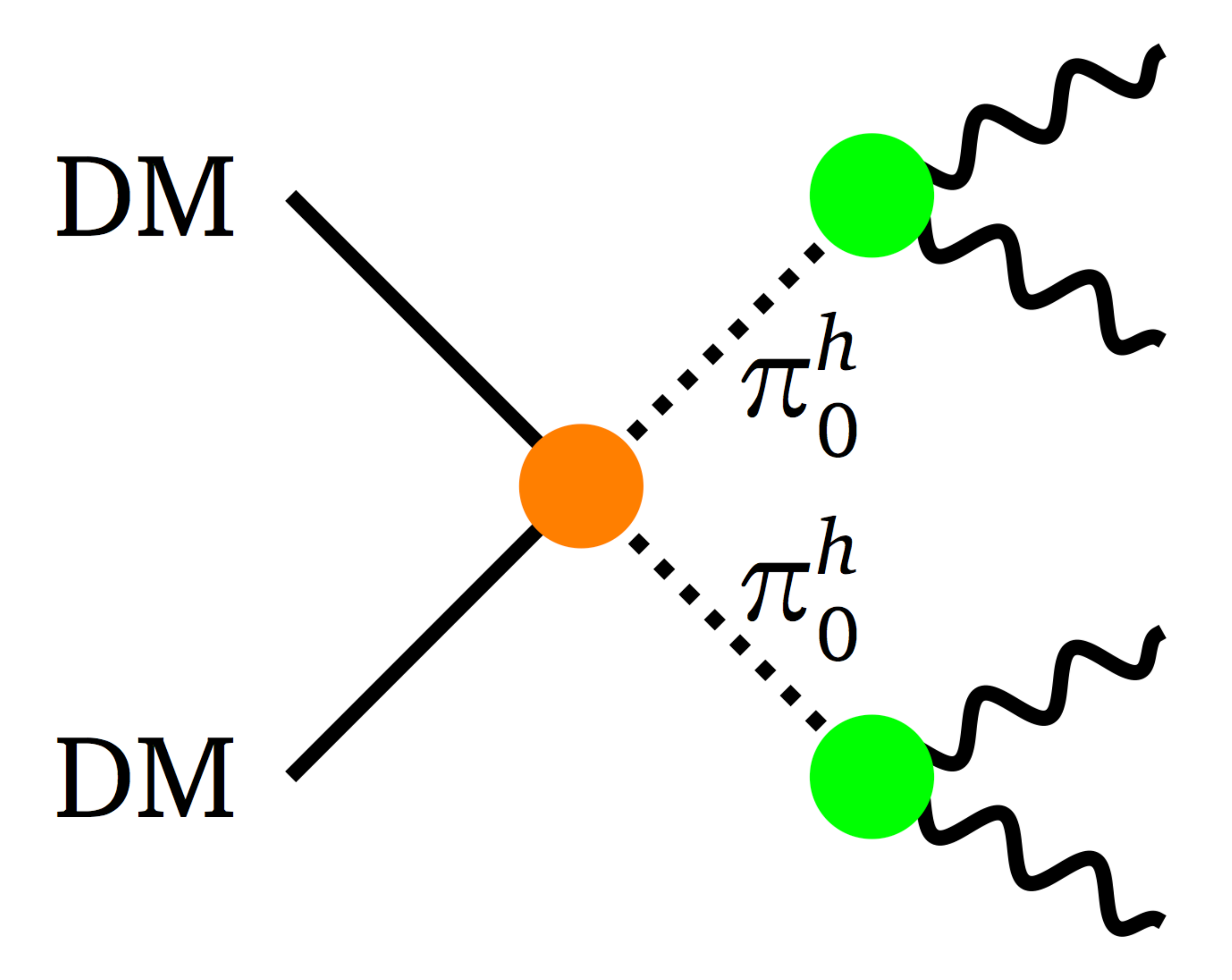}
\caption{The process ${\rm DM}+{\rm DM} \to \pi^h_0 + \pi^h_0 $, for a pseudoscalar $\pi^h_0$ which subsequently decays to photons, leads to a box-shaped gamma-ray spectrum~\cite{Fortin:2009rq,Ibarra:2012dw}. The goal of our model is to explain the narrowness of the box by placing the DM and $\pi^h_0$ in the same multiplet due to some symmetry, with nearly the same mass.\label{fig:boxtopology}}
\end{figure}

A strikingly different option is the possibility that the gamma ray lines are actually narrow box-shaped features~\cite{Fortin:2009rq,Ibarra:2012dw}. This occurs when dark matter annihilates to (pseudo)scalar states which in turn {\em decay} to two photons (or $\gamma+Z$), as shown in Figure~\ref{fig:boxtopology}. The gamma rays arising from these decays have energy bounded between $\frac{1}{2} \left(m_{\rm DM} \pm \sqrt{m_{\rm DM}^2 - m_\pi^2}\right)$, becoming a sharp line in the limit $m_\pi \to m_{\rm DM}$. This motivates the study of models with $m_{\rm DM} \approx 2 E_{\rm line} \sim 260$ GeV, with a pseudoscalar nearby in mass~\cite{Ibarra:2012dw,Buckley:2012ws,Chu:2012qy,Bai:2012new}. Because the annihilation process in this case is tree-level, it is much easier to accommodate the values of $\sigma v$ that fit the data. On the other hand, because the decay process, which is the only connection to the Standard Model, can be relatively delayed, one can imagine hidden sector dark matter that is difficult to probe in collider or direct detection experiments.

Our goal in this paper is simply to point out that the coincidence $m_{\rm DM} \approx m_{\pi}$ has a beautifully simple explanation if dark matter and the pseudoscalar $\pi$ are members of a multiplet. For example, consider low-energy QCD, in which the charged pions are slightly heavier than the neutral pion. In a world without weak interactions, the charged pions would be stable, but could annihilate to neutral pions. This will be the basis for our model: a heavier copy of QCD, with stable ``charged'' pions constituting the dark matter, where the charge is under a new U(1) symmetry. The neutral pion, through a higher-dimension operator, can decay to photons. This decay, in our model, gives rise to the gamma rays observed by Fermi-LAT. We will also point out that Big-Bang nucleosynthesis (BBN) constraints on the $\pi$ lifetime impose a limit to how weakly coupled dark matter and the Standard Model can be in such models.

{\bf A Simple Pion Model:} Our model for the narrow box-shaped gamma ray features mimics a subset of the fields of QCD. We take an SU($N$) gauge group with matter content displayed in Table~\ref{tab:fields}. U(1)$_X$ is a new abelian symmetry which stabilizes the dark matter. The $p$ and $q$ fields may be thought of as analogues of the up and down quark in QCD. In addition, we assume the existence of a light axion field $a$ which couples to the field strengths of both hypercharge and the SU($N$) group (denoted $H^a_{\mu\nu}$):
\beq
{\cal L}_{\rm axion} = \frac{c_B\alpha_Y}{8\pi} \frac{a}{f_a} B^{\mu\nu}{\tilde B}_{\mu\nu} + \frac{\alpha_N}{8\pi} \frac{a}{f_a} H^{a\mu\nu}{\tilde H}^a_{\mu\nu}.
\eeq
This axion gets a mass from SU($N$) instantons and is not the QCD axion.

\begin{table}[!h]
\begin{tabular}{c c c c}
& SU($N$) & U(1)$_X$  \\
\hline
$p$ & $\Box$ & $+1/2$ \\
${\overline p}$ & ${\overline \Box}$ & $-1/2$  \\
$q$ & $\Box$ & $-1/2$  \\
${\overline q}$ & ${\overline \Box}$ & $+1/2$ 
\end{tabular}
\caption{Field content of the model's hidden sector. The fields are all taken to be left-handed Weyl fermions.}
\label{tab:fields}
\end{table}

\begin{figure}[!h]
\includegraphics[width=0.7\columnwidth]{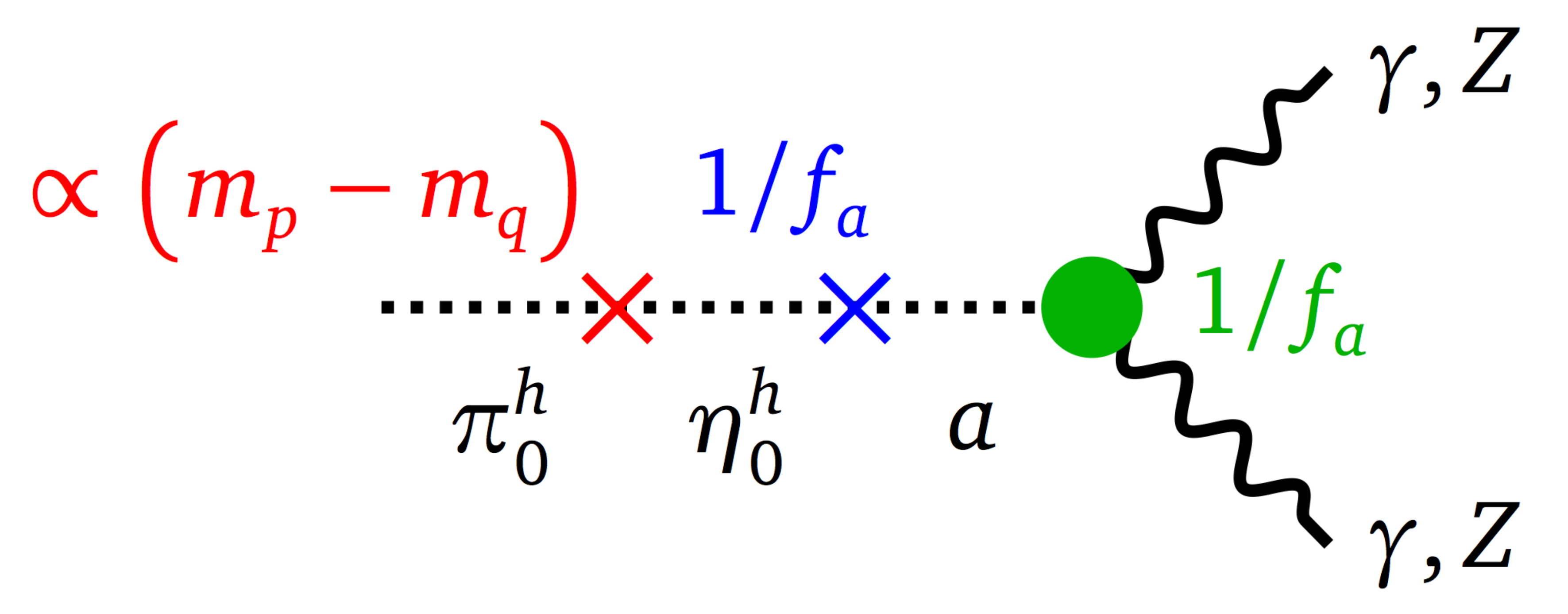}
\caption{The process $\pi^h_0 \to \gamma \gamma$ in our model. It proceeds by mixing with a light axion $a$ that couples both to $F{\tilde F}$ and $H{\tilde H}$. The $\pi^h_0$ can only decay in this way through its mixing with the $\eta_0^h$ state, which requires isospin breaking $m_p \neq m_q$. \label{fig:pi0decay}}
\end{figure}

We assume that there are mass terms $m _p {\overline p}p+m_q {\overline q} q$, with $m_p \neq m_q$ and $m_{p,q} < \Lambda_N$, where $\Lambda_N$ is the confinement scale of SU($N$). The theory above the scale $\Lambda_N$ enjoys a U(2)$_L \times$ U(2)$_R$ symmetry, which as in QCD is broken to the diagonal. This results in Nambu-Goldstone bosons $\pi^h_\pm, \pi^h_0,$ and $\eta_0^h$, where subscripts refer to U(1)$_X$ charges and the superscript $h$ reminds us that these are hidden-sector fields, not QCD pions. Like the $\eta'$ in QCD, the $\eta_0^h$ is not a true Nambu-Goldstone boson but obtains a mass through the U(1)$_A$ axial anomaly. Unlike QCD pions, $\pi^h_\pm$ are stable, due to being the lightest particles charged under U(1)$_X$.
We assume $\pi^h_\pm$ are dark matter, and the observed annihilation process is:
\beq
\pi^h_+ \pi^h_- \to \pi^h_0 \pi^h_0,~~~\pi^h_0 \to \gamma\gamma,~\gamma Z.
\label{eq:signaltopology}
\eeq
The $\pi^h_0$ decay proceeds through mixing as shown in Fig.~\ref{fig:pi0decay} and explained in more detail below. The $\pi^h$ fields are made massive by the explicit symmetry breaking $m_{p,q}$, and isospin breaking $m_p \neq m_q$ breaks all remaining symmetries except U(1)$_X$, allowing the $\pi^h_0$ and $\eta_0^h$ to mix. Because both the overall mass scale $m_{\pi_\pm}$ and the splitting $\delta m_\pi \equiv m_{\pi_\pm} - m_{\pi_0}$ are important for understanding the dark matter annihilation signal, we will briefly review the derivation of these quantities from the chiral Lagrangian (see e.g.~\cite{Manohar:1996cq} for details). We work with a nonlinear sigma model field $U = e^{i \pi^h/f_\pi}$, where
\beq
\pi^h = \left(\begin{matrix} \pi^h_0 + \eta_0^h & \sqrt{2} \pi^h_+ \\ \sqrt{2} \pi^h_- & -\pi^h_0 + \eta_0^h \end{matrix}\right).
\eeq
Taking $M$ to be a diagonal mass matrix for the fields $p$ and $q$, we can understand the masses and mixings of various states from the chiral Lagrangian ${\cal L} = \frac{1}{4} f_\pi^2 {\rm Tr}\left[\partial_\mu U^\dagger \partial^\mu U\right] + \mu \frac{f_\pi^{2}}{2} {\rm Tr}\left[U^\dagger M + M^\dagger U\right] -\frac{1}{2} m_{\eta_0}^2 \left(\eta_0^h\right)^2$. Here $\mu$ is determined by the GOR relation~\cite{GellMann:1968rz} to be $-\left<{\bar q}q\right>_0/f_\pi^2$ and the mass term for the $\eta_0^h$ represents the topological susceptibility effect~\cite{Witten:1979vv,Veneziano:1979ec}. This leads to a mass matrix in the $(\pi^h_0, \eta^h_0)$ basis:
\beq
M_0^2 =  \left(\begin{matrix} (m_p + m_q)\mu & (m_p - m_q)\mu \\ (m_p - m_q)\mu & (m_p+m_q)\mu + m_{\eta_0}^2 \end{matrix}\right).
\eeq
Assuming $m_{\eta_0}^2 \gg m_{p,q} \mu$, and continuing to expand the chiral Lagrangian, we derive a small splitting between the charged and neutral pion states:
\beq
m_{\pi_\pm}^2 & = & (m_p + m_q) \mu \\
\delta m_\pi & = & \frac{\left(m_p-m_q\right)^2 \mu^2}{2 m_{\pi_\pm} m_{\eta_0}^2}.\label{eq:splitting}
\eeq
Because the splitting is of second order in the quark mass difference, it is natural for the pion multiplet to be fairly degenerate. Of course, this mixing effect also means that the light mass eigenstate is not purely $\pi^h_0$, but contains an admixture $\frac{(m_q - m_p)\mu}{m_{\eta_0}^2}$ of the $\eta_0^h$ state.

The chiral Lagrangian leads to a scattering amplitude ${\cal A}(\pi^h_+ \pi^h_- \to \pi^h_0 \pi^h_0) = s/f_\pi^2 = 4 m_{\pi_+}^2/f_\pi^2$ at tree level, where in the last step we took the nonrelativistic limit relevant for dark matter annihilation. This implies that 
\beq
\sigma v & = & \frac{m_{\pi_+}^2}{4 \pi f_\pi^4}\sqrt{1 - \frac{m_{\pi_0}^2}{m_{\pi_+}^2}} \label{eq:sigmav} \\
& \approx & \frac{m_{\pi_+}}{4\pi f_\pi^4} \frac{\left|m_p - m_q\right| \mu}{m_{\eta_0}}. 
\eeq
To relate some of the parameters appearing in this formula, we will scale up QCD with the large-$N$ estimates $\mu \approx 77f_\pi/\sqrt{N}$ and $m_{\eta_0} \approx 31 f_\pi/N$. 

Using the large-$N$ estimate for the matrix element $\left<0\left|H^{a\mu\nu}{\tilde H}^a_{\mu\nu}\right|\eta_0\right>$~\cite{Witten:1979vv}, we can estimate that the mass mixing between the light mostly-$\pi^h_0$ mass eigenstate and the axion $a$ is:
\beq
{\cal L}_{\rm mix} \approx \frac{1}{2\sqrt{2}} \frac{\left(m_p - m_q\right)\mu f_\pi}{f_a} a \pi^h_0.
\eeq 
This mixing leads to a positive shift in $m_{\pi_0}$ at the second order in perturbation theory. Requiring it to be smaller than the negative contribution in Eq.~(\ref{eq:splitting}), we find $f_a \simgt f_\pi m_\eta/m_\pi$. For a TeV scale QCD-like sector, this amounts to $f_a \simgt 10^4$ GeV.
The $\pi^h_0$ decay width is:
\beq
\Gamma(\pi^h_0 \to \gamma\gamma) = \frac{c_B^2 \alpha^2  }{2048 \pi^3} \left(\frac{m_p-m_q}{m_p+m_q}\right)^2 \frac{f_\pi^2 m_{\pi_0}^3}{f_a^4}.
\eeq
The relative widths of the subdominant processes are (where $t_W \equiv \tan\theta_W$ with $\theta_W$ the Weinberg angle):
\beq
\frac{\Gamma(\pi^h_0 \to \gamma Z)}{\Gamma(\pi^h_0\to\gamma\gamma)} &= &2 t_W^2 \left(1 - \left(\frac{m_Z}{m_{\pi_0}}\right)^2\right)^3 \approx 0.4,\\
\frac{\Gamma(\pi^h_0 \to Z Z)}{\Gamma(\pi^h_0\to\gamma\gamma)} &= & t_W^4 \left(1 - \left(\frac{2 m_Z}{m_{\pi_0}}\right)^2\right)^{3/2} \approx 0.03.
\eeq
Axion couplings to $W{\tilde W}$ and $G{\tilde G}$ which change the branching ratios are allowed, provided they are small enough to evade continuum bounds. Almost independent of the cosmological history of the universe, we expect that the relic abundance of $\pi^h_0$ and $\pi^h_\pm$ states would be comparable, because they are related by a symmetry that is only mildly broken. (This is generic but loopholes could exist, e.g. a chemical potential for U(1)$_X$.) Given the present-day relic abundance of $\pi^h_\pm$, the hadronic decays of the $Z$ lead to a BBN constraint that the $\pi^h_0$ lifetime be $\simlt 100$ seconds~\cite{Jedamzik:2006xz}, imposing $f_a \simlt 10^7$ GeV. Note a similar bound will apply in {\em any} model in which DM and $\pi_0^h$ abundances are related by a symmetry.

The axion mass is estimated as
\beq
m_a^2 \approx \frac{1}{f_a^2} \frac{m_p m_q}{\left(m_p + m_q\right)^2} m_{\pi_0}^2 f_\pi^2,
\eeq
up to subleading corrections proportional to $\left(m_p - m_q\right)^2$~\cite{Kim:2008hd}. Its decay width is
\beq
\Gamma(a \to \gamma\gamma) = \frac{c_B^2 \alpha^2}{256\pi^3} \frac{m_a^3}{f_a^2}.
\eeq
The axion-like particles' cosmological and laboratory constraints are summarized in~\cite{Cadamuro:2011fd, Hewett:2012ns}. In our case, they restrict  $f_a \simlt 10^6$ GeV. 

Taking into account of all the constraints on the hidden pions and axion and one additional cosmological constraint $N \simlt 12$~\cite{Kaplan:2006yi}, we give a benchmark point
\beq
f_\pi &=& 0.8\, {\rm TeV},\,N =3,\,m_p=1.53 \,{\rm GeV},\,m_q= 0.31 \,{\rm GeV}, \nonumber \\
 \delta m_\pi&=& 0.05\,{\rm GeV},\,\sigma v_{\rm eff} = 1.3 \times 10^{-27} {\rm cm}^3/{\rm s}, \nonumber \\
m_{\pi_\pm} &=& 256\, {\rm GeV},\,f_a = 10^5\,{\rm GeV}, \,m_a =0.6\,{\rm GeV},\nonumber \\
c_B&=&1,\, \tau_{\pi^h_0}=1 \times 10^{-8}\,{\rm s},\,\tau_{a}=5 \times 10^{-6}\,{\rm s}. 
\eeq
Here $\sigma v_{\rm eff}$ is the cross section from Eq.~\ref{eq:sigmav} weighted for direct comparison to the line strength for the hypothesis ${\rm DM}~{\rm DM} \to \gamma\gamma$ as estimated in the literature. (In other words, we have rescaled by a factor of 2 because we have twice as many photons per dark matter annihilation.) Further possible bounds on the self-interactions $\pi^h_+ \pi^h_- \to \pi^h_+ \pi^h_-$~\cite{Markevitch:2003at,MiraldaEscude:2000qt} are far too weak to constrain this model. To see this, note that these interactions are controlled by the chiral Lagrangian and have $\sigma v \sim m_{\pi_+}^2/(4\pi f_\pi^4) \approx 5~{\rm pb}$, many orders of magnitude smaller than the $\sim 0.1~{\rm barn}/{\rm GeV}$ bounds on self-interactions.

As an alternative to the axion with very similar phenomenology, one could add heavy fields $\ell, {\bar \ell}$ charged under both SU($N$) and U(1)$_Y$. Integrating them out produces operators coupling the $\eta_0^h$ to $B_{\mu \nu}{\tilde B}^{\mu\nu}$, suppressed by $m_\ell^4$. Again, the BBN constraint that the $\tau_{\pi^h_0} \simlt 100$ s imposes an upper bound, in this case $m_\ell \simlt 10^5$ GeV.

{\bf Width of the box-like feature:} In our benchmark point, the width of the box-like spectral feature is
\beq
\Delta E = \sqrt{2 m_{\pi_+} \delta m} \approx 2.5\sqrt{N} \left|m_p - m_q\right|  \approx 5~{\rm GeV}.
\eeq
One can ask what range of widths for the feature are expected in a range of models, if we fix the cross section and dark matter mass. The cross section scales as $\sigma v \sim m_{\pi_+}  \Delta E/f_\pi^4$, so to fix $\sigma v$ one must scale $f_\pi \sim \Delta E^{1/4}$. (The reasonable range of $N$ is limited, so varying $N$ as well as $f_\pi$ has at most an order-one effect.) On the other hand, the pion mass scales as $\sim (m_p + m_q)f_\pi$. Hence, if we aim to fix the dark matter mass and cross section, we can only significantly alter $\Delta E$ by {\em tuning} the ratio $\left|m_p - m_q\right|/(m_p + m_q)$. A box much wider than 10 GeV is disfavored by data. A box of width 100 MeV corresponds to about a 1\% tuning of the quark masses, and narrower boxes are even more tuned. Hence, the range of spectral features to consider is from several hundred MeV to of order 10 GeV.

{\bf Relic abundance:} The light hidden axions will connect the hidden QCD sector and the SM in the same thermal bath at high temperatures in the early Universe. However, $\sigma v(\pi^h_+ \pi^h_- \to \pi^h_0 a)$ is so small that the axion will first freeze out from the hidden sector, yielding an overly large thermal relic abundance for $\pi^h_{\pm}$. One way to avoid this difficulty is through a nonthermal cosmology with late entropy production, which can dramatically change the dependence of dark matter abundance on the annihilation cross section~\cite{Chung:1998rq, Giudice:2000ex}. This will allow the right relic abundance to be obtained, for instance for certain lifetimes of late-decaying particles.

{\bf Other possibilities:} Although we have discussed a particular model, it bears repeating that the central idea is that dark matter is a stable component of a multiplet that also contains a particle that can decay to photons, and that the states in this multiplet are nearly degenerate. Many other models could realize this paradigm. For example, consider a set of states related by supersymmetry, with the dark matter a fermionic state ${\tilde X}$ protected by $R$-parity with a decaying scalar superpartner $X \to \gamma\gamma$. Such nearly degenerate boson/fermion pairs are easily accommodated in simple models with a light gravitino~\cite{Fan:2011yu}, but these are bad dark matter candidates because the fermion will decay to its superpartner and a gravitino. In the case of heavy gravitinos, sequestering the multiplet from large supersymmetry breaking complicates the model~\cite{Fan:2012jf}. An intermediate regime, with gravitino mass $m_{\tilde G} \simgt m_{\tilde X} - m_X \sim 100~{\rm MeV}$, may form a good compromise. Nor are pions and supermultiplets the only options. Existing models of decaying dark matter for the line~\cite{Kyae:2012vi,Park:2012xq} could be re-engineered as models for the decay of a state that dark matter annihilates into.

{\bf Conclusion:} Typically, observable indirect detection signals of dark matter are thought of as pointing to new particle physics near the weak scale, interacting with the Standard Model through renormalizable interactions. Here we have shown that the interactions responsible for an indirect detection signal can be weaker, suppressed by high-dimension operators. The mass degeneracy needed for the topology (\ref{eq:signaltopology}) to explain a line-like signal arises from a symmetry. This symmetry in turn predicts similar relic abundance for $\pi^h_\pm$ and $\pi^h_0$, so BBN poses an interesting constraint on the lifetime $\tau_{\pi^h_0} \simlt 100~{\rm s}$. Thus, the dark matter can be in a hidden sector, but it cannot be completely hidden, and the possibility of probing such sectors in terrestrial experiments deserves more attention. The coincidence of the dark matter mass with the weak scale is a puzzle, and it is tempting to suggest that both scales could arise from a common origin in supersymmetry breaking. But this model stands logically apart from such considerations; it is motivated by data alone, and confirmation of the gamma ray lines or correlated signals by future experiments~\cite{Bergstrom:2012vd,Laha:2012fg} would be the first necessary test of the idea. Our benchmark point gives a box of 5 GeV width rather than a line. This is narrow relative to the Fermi-LAT energy resolution, but a confirmation that the shape is box-like rather than line-like would be the smoking gun for a scenario like ours.

{\bf Acknowledgments:} We thank David Krohn and Itay Yavin for early-stage collaboration. We also thank Meng Su and Neal Weiner for useful comments. JF, MR are supported in part by the Fundamental Laws Initiative of the Harvard Center for the Fundamental Laws of Nature. JF acknowledges the hospitality of the Aspen Center for Physics, which is supported by the National Science Foundation Grant No. PHY-1066293.

\end{document}